\documentclass[twocolumn]{aastex631}

\usepackage{makecell}
\usepackage{comment}
\usepackage{natbib}
\usepackage{multirow}
\usepackage{booktabs}
\usepackage{epsfig}
\usepackage{ulem}
\usepackage{rotating}
\usepackage{graphicx}
\usepackage{url}
\usepackage{xcolor}
\usepackage{amsmath}
\usepackage{color}
\usepackage{savesym}
\savesymbol{tablenum}
\usepackage{siunitx}
\usepackage[normalem]{ulem}
\usepackage{appendix}
\usepackage{enumerate}
\usepackage{gensymb}
\usepackage{enumitem}

\usepackage{longtable}
\usepackage{afterpage}
\usepackage{threeparttable}
\usepackage{tabularx}
\usepackage[bottom]{footmisc}

\DeclareSIUnit\parsec{pc}
\DeclareSIUnit\h{\textit{h}}

\begin{document}

    \title{Calibrating the Tully-Fisher Relation to Measure the Hubble Constant }
    
    \author[0000-0002-4934-5849]{Daniel Scolnic}
    \affiliation{Department of Physics, Duke University, Durham, NC 27708, USA}


   \author[0000-0002-1809-6325]{Paula Boubel}
    \affiliation{Research School of Astronomy and Astrophysics, The Australian National University, Mount Stromlo Observatory, Canberra, ACT 2611, Australia}
    
    \author[0000-0002-4934-5849]{Jakob Byrne}
    \affiliation{North Carolina School of Science and Math, Durham, NC 27708, USA}

    \author[0000-0002-6124-1196]{Adam G.~Riess}
    \affiliation{Space Telescope Science Institute, Baltimore, MD 21218, USA}
    \affiliation{Department of Physics and Astronomy, Johns Hopkins University, Baltimore, MD 21218, USA}

  \author[0000-0002-5259-2314]{Gagandeep S. Anand}
\affiliation{Space Telescope Science Institute, 3700 San Martin Drive, Baltimore, MD 21218, USA}

\begin{abstract}

\cite{Boubel24} recently used the Tully-Fisher (TF) relation to measure calibrated distances in the Hubble flow and found $H_0= 73.3 \pm 2.1 \textnormal{(stat)} \pm 3.5 \textnormal{(sys)}$ km/s/Mpc.  The large systematic uncertainty was the result of propagating the conflict between two sources of empirical distance calibration: a difference in zeropoint when calibrating the TF relation with either Type Ia supernovae (SNe Ia) or Cepheids and Tip-of-the-Red-Giant-Branch (TRGB) and an apparent difference in zeropoint between two distinct TRGB datasets.  We trace the SN Ia-based calibration used in the TF analysis to a study where $H_0$ was fixed to 70 km/s/Mpc rather than measured, (with host distances derived from redshifts and the Hubble law), thus introducing a discrepancy with the other empirically calibrated indicators. In addition, we trace the difference in TRGB zeropoints to a miscalibration of $0.14$ mag that should be $\sim0.01-0.04$ mag.  Using the consistent Cepheid and TRGB calibration from \cite{Boubel24} while removing the problematic data reduces the systematic error by a factor of two and results in $H_0 = 76.3 \pm 2.1 \textrm{(stat)} \pm 1.5 \textrm{(sys)}$ km/s/Mpc.  This measurement is consistent with previous determinations of $H_0$ using the TF relation.  We also show that most determinations of $H_0$ measurements that replace Type Ia supernovae measurements with another far-field distance indicator yield $H_0>73$ km/s/Mpc, reinforcing previous findings that the Hubble tension is not tied to any one distance indicator.

\end{abstract}
\keywords{}

\section{Introduction}
The discrepancy between direct measurements of the current expansion rate of the universe and inferred measurements from the early universe, commonly referred to as the `Hubble Tension,' remains a significant challenge in cosmology (see \citealp{Verde23} for a review). In the most widely used distance ladder approach \citep{Riess22,Burns18}, Type Ia supernovae (SNe Ia) are employed to measure the distance-redshift relation, enabling the calculation of the Hubble constant, $H_0$. A separate distance ladder method utilizes the Tully-Fisher (TF) relation, which links a spiral galaxy's rotational velocity to its luminosity, and can trace the distance-redshift relation. An overview of this approach and its application to measure $H_0$ is presented in \cite{Said23}, which compiles 28 published constraints on $H_0$ (e.g. \citealp{Sorce14,
Neil14, Schombert2020, Kourkchi20, Kourkchi22,Courtois23}).

Interestingly, in the last 35 years, every single $H_0$ measurement using the TF relation as the final rung of the distance ladder returned $H_0>70$. In this analysis, we look at the most recent TF measurement in \cite{Boubel24}, which finds $H_0= 73.3 \pm 2.1 \textnormal{(stat)} \pm 3.5 \textnormal{(sys)}$ km/s/Mpc.  As the systematic uncertainty is rather large, a better understanding of this result can help inform the overall robustness of this distance-ladder approach.

\begin{figure*}
    \centering
    \includegraphics[width=2\columnwidth]{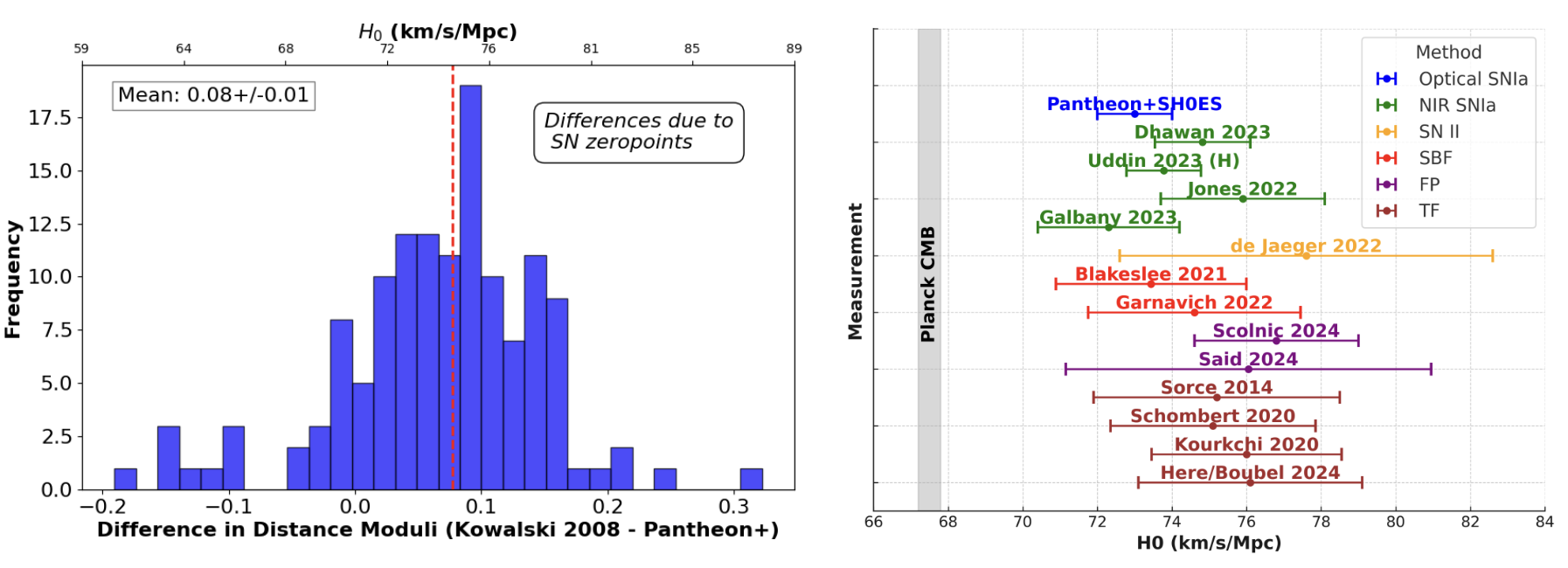}
\caption{(Left) The difference in supernova distance modulus values, as shown on the bottom x-axis, of SNe Ia from the catalog used in B24 and the set from Pantheon+ \cite{Scolnic22,Brout22}. The top x-axis shows the impact on $H_0$ in B24 if the sample used, previously calibrated to $H_0=70$, is replaced with one calibrated using Cepheids or TRGB like that in Pantheon+. (Right) A compilation of distance ladder measurements of $H_0$ in comparison to the Pantheon+SH0ES, where the third rung of the distance ladder is redone using various techniques. The legend shows the different techniques, which are described in Section~\ref{sec:compilation}.}
    \label{fig:whisker}
\end{figure*}

Whether using SNe Ia or the TF relation as the final rung of the distance ladder, both methods require their luminosities to be calibrated. B24 calibrates the zeropoint of their TF relation using Cepheids, the Tip-of-the-Red-Giant-Branch (TRGB), and SNe Ia. Variations in the zeropoint directly affect the measured value of $H_0$. B24 finds significant differences (on the order of $0.1$ mag or $\Delta H_0\sim3$) when using SNe Ia versus Cepheids/TRGB or when employing different TRGB analyses. These discrepancies contribute to a substantial systematic uncertainty in B24’s final estimate of $H_0$, which we examine in detail here.

The statistical uncertainty in a TF-based distance ladder is from the calibration and the distance-redshift relation rung.  The TF approach leverages spiral galaxies, which are the most numerous type of galaxies, and the Cosmicflows-4 sample \citep{Kourkchi20,Kourkchi22} has a sample across the sky of $\sim10,000$ galaxies.  Although the intrinsic scatter of the TF relation is large at around $20\%$ of the distances, this uncertainty will be small for the distance-redshift relation as the statistics are in the thousands. However, for the calibrating rung, in which the TF relation is calibrated by a primary distance indicator like Cepheids or TRGB, the sample is in the tens, and the uncertainty will be larger. 

As the systematic uncertainty found in B24 is much larger than the statistical uncertainty and the statistical uncertainty can be directly calculated from the dispersion of the calibration rung, we focus more on the systematic component of the measurement.  After doing so, we place this new measurement in the context of other TF measurements of $H_0$, as well as constraints from other techniques.

\section{Analysis}

The large systematic uncertainty in B24 can be traced to two discrepancies in calculating the zeropoint of the TF relation.  We discuss both below and then present a revised measurement after reconciling these discrepancies.

\subsection{A different zeropoint when using Cepheid/TRGB versus using SNe Ia}

The first discrepancy we investigate is a 0.1 mag difference in the zeropoint when using SNe Ia versus various primary distance indicators (e.g., Cepheids or TRGB). Unlike Cepheids and TRGB, SNe Ia are not directly calibrated by a geometric anchor and are themselves typically calibrated using Cepheids or TRGB, making a true discrepancy hard to understand. B24 employs the \cite{Makarov14} (hereafter M14) catalog of distances, which compiles measurements from stellar indicators. We note that none of these come from the SH0ES Cepheid measurements \citep{Riess22}.  For the SNe Ia, the distances are derived from \cite{Kowalski08} (hereafter K08). K08 does not explicitly address the assumption of $H_0$ in their analysis, as the $H_0$ term is degenerate with the absolute magnitude of SNe Ia when measuring cosmological parameters and is typically marginalized over in the context of dark energy studies.

We use the distances from K08 and crossmatch the supernovae with those in the Pantheon+ dataset \citep{Scolnic22}, where the distances are calibrated using Cepheid measurements \citep{Riess22}. There are 200 SNe in common between the datasets, and the distribution of differences is shown in Fig. 1a. We find a net offset of $0.08 \pm 0.02$ mag. Given that the $H_0$ measured in Pantheon+SH0ES is 73 km/s/Mpc, we infer that the $H_0$ assumed in K08 is likely 70 km/s/Mpc. This assumption is common in SNe Ia cosmological analyses of the same era \citep{Kessler09, Guy10, Scolnic18}, although the distances provided by K08 were not intended to directly measure $H_0$.  A conflict between distances derived from an assumed value of $H_0$ versus empirical measurements is expected and does not provide any meaningful measure of systematic uncertainty in calibration.

\subsection{A different zeropoint when using different sets of TRGB distances}

The second discrepancy arises from an apparent 0.14 mag difference in the zeropoint when using the TRGB sample in M14 versus the TRGB analysis from \cite{Freedman19} (hereafter F19). The TRGB distances from M14 are derived from \cite{Rizzi07} (hereafter R07), whereas the TRGB distances cited to be in \cite{Freedman19} were actually misidentified from another catalog and not directly compared with any TF distances.  
The difference in absolute magnitude of the TRGB between R07 and F19 is only 0.01 mag, which is much smaller than the claimed 0.14 mag difference.  A separate study by \cite{Anand22} recalibrated the measurement by R07 with new measurements of the anchor galaxy NGC 4258, and found a fainter absolute luminosity of $M_{F814W} \sim-4.0$.  This would increase $H_0$ in B24 for this calibration path by $\sim2$ km/s/Mpc. In order to make minimal alterations to the B24 analysis, we retain the TRGB calibration from R07 and leave the \cite{Anand22} recalibration as a systematic uncertainty.


\subsection{Revising B24}

A straightforward solution is to discard the SNe Ia and misidentified TRGB distances and rely solely on the direct distances from Cepheids and TRGB in M14. B24 reports $H_0$ measurements for this case: $H_0 = 75.7 \pm 3.8$ km/s/Mpc when using SDSS i-band TF measurements, and $H_0 = 76.3 \pm 2.1$ km/s/Mpc when using WISE W1-band measurements. Given the larger number of calibrator galaxies from the WISE W1-band measurement, we proceed with WISE from here on. The uncertainties in this $H_0$ measurement are only statistical, with systematic uncertainties arising from the large zeropoint calibration differences. After simplifying the analysis to use only Cepheids and TRGB, the systematic uncertainty from the difference between these two calibrations is reduced to 0.5 km/s/Mpc. The result from Cepheids is $76.1\pm3.0$ and the result from TRGB is $76.1 \pm 2.7$. In B24, the dispersion in recovered $H_0$ values using various techniques to determine the TF zeropoint was used to set the large systematic uncertainty of 3.5 km/s/Mpc.  With this curated set, the dispersion is much smaller, so it is likely that we would underestimate the actual uncertainty.

One source of uncertainty that must be included is the geometric distance uncertainty of NGC 4258, which is 0.032 mag. As discussed above, there is also uncertainty in the TRGB zeropoint calibration, as large as a 0.05 mag shift that would produce a higher value of $H_0$ for TRGB.  In M14, they show that previous determinations of TRGB zeropoints vary on the 0.02-0.03 mag level.  As we combine the Cepheid and TRGB calibrations, and given the consistent TF-relation calibration from each before any modification, we attribute 0.025 mag to the zeropoint uncertainty.  Combining the geometric uncertainty and zeropoint uncertainty, the total systematic uncertainty is $\sqrt{0.032^2+0.025^2}=0.04$ mag. Our final measurement therefore is $H_0 = 76.3 \pm 2.1 \textrm{(stat)} \pm 1.5 \textrm{(sys)}$ km/s/Mpc when using WISE W1-band measurements. We choose WISE for our final constraint as there is a larger number of calibrating galaxies for this sample.  Our final result aligns with the other main TF-based distance ladder measurement as discussed below.

\section{Compiling $H_0$ Measurements with Different Distance-Redshift Relations}
\label{sec:compilation}

An interesting finding is that with these changes to the zeropoints used in B24, we can directly compare the inferred value of $H_0$ when using SNe Ia as the last rung of the distance ladder versus the TF relation.  This comparison is self-consistent, as the Cepheids / TRGB used to calibrate SNe Ia as in \cite{Anand22,Riess22,Scolnic22} are the same.  With these zeropoint calibrations, the inferred value of $H_0$ is $\sim72-74$ km/s/Mpc.  This is $\Delta H_0 \sim 3$ km/s/Mpc smaller than when using the TF relation.

To pursue this further, we compile other $H_0$ measurements that use Cepheid calibrations, but vary the distance method in the distance-redshift relation.  This is shown in Fig. 1b.  These include:
 \begin{enumerate}[itemsep=1pt, topsep=2pt]
 \item Optical SNe Ia light curves: $73\pm1$ \citep{Riess22}.
\item The Tully Fisher relation: 
$75.2\pm3.3$ \citep{Sorce14}
$75.1	\pm2.75$ \citep{schombert20}; $76.0\pm2.55$ \citep{Kourkchi20}; $76.3 \pm 2.16$ km/s/Mpc (Here/B24).  We only include measurements with systematic uncertainties propagated.
\item The Fundamental Plane relation: $76.5\pm2.2$ \citep{Scolnic24}; $76.05\pm4.90$ \citep{Said24}.
\item SNe II: $77.6\pm5$ \citep{dejaeger20}
\item SNe Ia NIR: $75.8\pm 5.73$ \citep{galbany23}, ($74.82\pm1.28$) \citep{Dhawan23}, $75.9\pm2.2$ \citep{Jones22}, $73.78\pm1$ \citep{Uddin23}
\item SBF: $73.44\pm2.55$ \citep{Blakeslee21}; $74.6\pm2.85$ \citep{Garnavich23}.
\end{enumerate}

 Interestingly, we find that the usage of SNe Ia yields one of the {\it lowest} values of the Hubble constant, and most measurements yield higher values with a range of $73-77$ km/s/Mpc.  \cite{Tully23} find a similar result that the calibration is lowered by 2.5 km/s/Mpc if the only coupling is through SN Ia. It is unclear if the difference in $H_0$ between SN Ia and other far-field indicators is significant, although interesting in the context of the Hubble tension that replacement of SNIa with any method makes the tension greater.

\section{Conclusion}

We correct two discrepancies in the zeropoint calibration of the TF relation used to measure $H_0$, which increases the recovered value of $H_0$ and reduces the systematic uncertainty.  We find $H_0 = 76.3 \pm 2.1 \textrm{(stat)} \pm 1.5 \textrm{(sys)}$ km/s/Mpc.  This value is higher than the typical values of $H_0$ recovered using SNe Ia as the last rung of the distance ladder.  We find that when replacing SNe Ia as the last rung, the measured values of $H_0$ are typically on the higher side.  This offers no respite for the lingering Hubble Tension.

\bibliographystyle{mn2e}
\bibliography{main}{}

\end{document}